\documentclass[12pt]{iopart}

\usepackage{graphicx}  
\begin{document}

\title[]{New limits on the $\beta^{+}$EC 
and ECEC processes in $^{120}$Te}

\author{A S Barabash$^1$, F Hubert$^2$, Ph Hubert$^2$ and V Umatov$^1$}

\address{$^1$ Institute of Theoretical and Experimental Physics, 
B. Cheremushkinskaya 25, 117218 Moscow, Russia}
\address{$^2$ Centre d'Etudes Nucl\'eaires,
IN2P3-CNRS et Universit\'e de Bordeaux, 33170 Gradignan, France}
\ead{barabash@itep.ru}
\begin{abstract}
New limits on the double beta processes for $^{120}$Te have been obtained using a 
400 cm$^3$ HPGe detector and a source consisting of natural 
Te0$_2$ powder. At a confidence level of 90\% the limits are $0.19\times 10^{18}$ y 
for the $\beta^+$EC$(0\nu + 2\nu)$  transition to the ground state, 
$0.75\times 10^{18}$ y for the ECEC$(0\nu + 2\nu)$ transition to the first 
2$^+$ excited state of $^{120}$Sn (1171.26 keV) and $(0.19-0.6)\times 10^{18}$ y 
for different ECEC($0\nu$) captures to the ground state of $^{120}$Sn.

\end{abstract}

\maketitle

\section{Introduction}
Most double beta decay investigations have concentrated on the $\beta^-\beta^-$ decay 
(see recent reviews \cite{BAR06,BAR06a,CRE06}). Much less attention has been given to the investigation of 
$\beta^+\beta^+$, $\beta^+$EC and ECEC processes. Here EC denotes the electron capture process. 
Although these alternate processes have been studied from time to time in the past (see review [4] and 
compilation of existing data before April 2001 in \cite{TRE02}) the field should see further exploration. 

There are 34 candidates for these processes. 
Only 6 nuclei can undergo all the above mentioned processes and 16 nuclei can undergo $\beta^+$EC 
and ECEC while 12 can undergo only ECEC. Detection of the neutrinoless mode in the 
above processes enable one to determine the effective Majorana neutrino mass 
$\left<m_\nu\right>$, parameters of right-handed current admixture in electroweak 
interaction ($\left<\lambda\right>$ and $\left<\eta\right>$), etc. Detection of the 
two-neutrino mode in the above processes lets one determine the magnitude of 
the nuclear matrix elements involved, which is very important in view of the theoretical 
calculations for both the $2\nu$ and the $0\nu$ modes of double beta decay. Interestingly, 
it was demonstrated in Ref. \cite{HIR94} that if the 
$\beta^-\beta^-(0\nu)$ decay is detected, then the experimental limits on the 
$\beta^+EC(0\nu)$ half-lives can be used to obtain information about the relative 
importance of the Majorana neutrino mass and right-handed current admixtures in 
electroweak interactions.

The $\beta^+\beta^+$ and $\beta^+$EC processes are less favorable due to 
smaller kinetic energy available for the emitted particles and
Coulomb barrier for the positrons. However, an 
attractive feature of these processes from the experimental point of view is the possibility of detecting 
either the coincidence signals from four (two) annihilation $\gamma$-rays and two (one) 
positrons, or the annihilation $\gamma$-rays. It is 
difficult to investigate the ECEC process because one only detects the low energy X-rays. It is 
interesting to search for transitions to the excited states of daughter nuclei, 
which are easier to detect given the cascade of higher energy gammas \cite{BAR94}. One can foresee 
that with improvement in the experimental techniques and increasing the mass of the detectors, 
the search for $\beta^+\beta^+$, $\beta^+$EC and ECEC processes will be 
considerably extended.

During the last few years, interest in the $\beta^+\beta^+$, $\beta^+$EC and ECEC processes 
has greatly increased. For the first time a positive result was obtained in a geochemical 
experiment with $^{130}$Ba, where the ECEC$(2\nu)$ process was detected with a 
half-life of $(2.2 \pm 0.5)\times 10^{21}$ y \cite{MES01}. Recently new  
limits on the ECEC($2\nu$) process in promising candidate isotopes ($^{78}$Kr and $^{106}$Cd) 
were established 
($1.5\times 10^{21}$ y and 
$6.2\times 10^{19}$ y, respectively \cite{GAV06,RUK06}). Also, 
for the first time 
limits on the double beta decay processes ($\beta^+$EC and ECEC) in $^{74}$Se have 
been obtained \cite{BAR06b}. Attention here was paid to the ECEC($0\nu$) transition to 
the $2^+_2$ excited state of $^{74}$Ge (1204.2 keV). In this decay there is a possible 
enhancement of the decay rate by several order of magnitude, because 
the ECEC($0\nu$) process is nearly degenerate with an excited state in the daughter 
nuclide. Under resonant conditions such ECEC($0\nu$) processes can have the same sensitivity 
to neutrino mass as $\beta^-\beta^-(0\nu)$ decay \cite{BAR06b,BER83}. In Ref. \cite{BEL03} 
there are limits on the ECEC($2\nu$) process in $^{136}$Ce and $^{138}$Ce at the 
level $\sim$ 10$^{16}$ y. Different $\beta^+\beta^+$, $\beta^+$EC and ECEC processes in $^{106}$Cd,
$^{108}$Cd and $^{180}$W have been investigated with a sensitivity of $\sim$ 10$^{17}$-10$^{19}$ y 
\cite{DAN03}.
Very recently $\beta^+\beta^+$, 
$\beta^+$EC and ECEC processes in $^{120}$Te, $^{106}$Cd, $^{108}$Cd and $^{64}$Zn 
were investigated in the framework of COBRA test measurements which had a sensitivity on the level of 
$\sim$ 10$^{15}$-10$^{17}$ y \cite{KIE03,WIL06}. Among the recent papers are a few new theoretical 
papers with half-life estimations \cite{SUH03,DOM05,CHA05,RAI06,SHU07}. Nevertheless the 
$\beta^+\beta^+$, $\beta^+$EC and ECEC processes have  not been investigated very well, both theoretically 
and experimentally. One can imagine some unexpected results here, which is why any 
improvements in experimental sensitivity for such transitions has merit. 

In this article the results of an experimental investigation of the $\beta^+$EC and 
ECEC  processes in $^{120}$Te (transitions to the ground and excited states of $^{120}$Sn) 
are presented. 
The decay scheme for the triplet $^{120}$Sn-$^{120}$Sb-$^{120}$Te is shown in Fig. 1. 
The Q value of the transition is $(1700.1 \pm 10.3)$ keV \cite{AUD03} and natural abundance 
of $^{120}$Te is 0.089\%. 
The following processes are possible:

\begin{equation}
e^-_b + (A,Z) \rightarrow (A,Z-2) + e^{+} + X   \hspace{2cm}  (\beta^+EC; 0\nu)   
\end{equation}

\begin{equation}
e^-_b + (A,Z) \rightarrow (A,Z-2) + e^{+} + 2\nu + X   \hspace{1cm}  (\beta^+EC; 2\nu)   
\end{equation}

\begin{equation}
2e^-_b + (A,Z) \rightarrow (A,Z-2) + 2X   \hspace{2.5cm}  (ECEC; 0\nu)   
\end{equation}

\begin{equation}
2e^-_b + (A,Z) \rightarrow (A,Z-2) + 2\nu + 2X   \hspace{1.5cm}  (ECEC; 2\nu)   
\end{equation}
where e$_b$ is an atomic electron and X represents an X-rays or Auger electrons.

In the case of ECEC process there is a possible transition to the 2$^+$ excited state (1171.26 keV).
Reported here is the search for the $\beta^+$EC and ECEC transitions using a germanium 
detector to look for $\gamma$-ray lines corresponding to the various decay schemes.

\begin{figure*}
\begin{center}
\includegraphics[width=12cm]{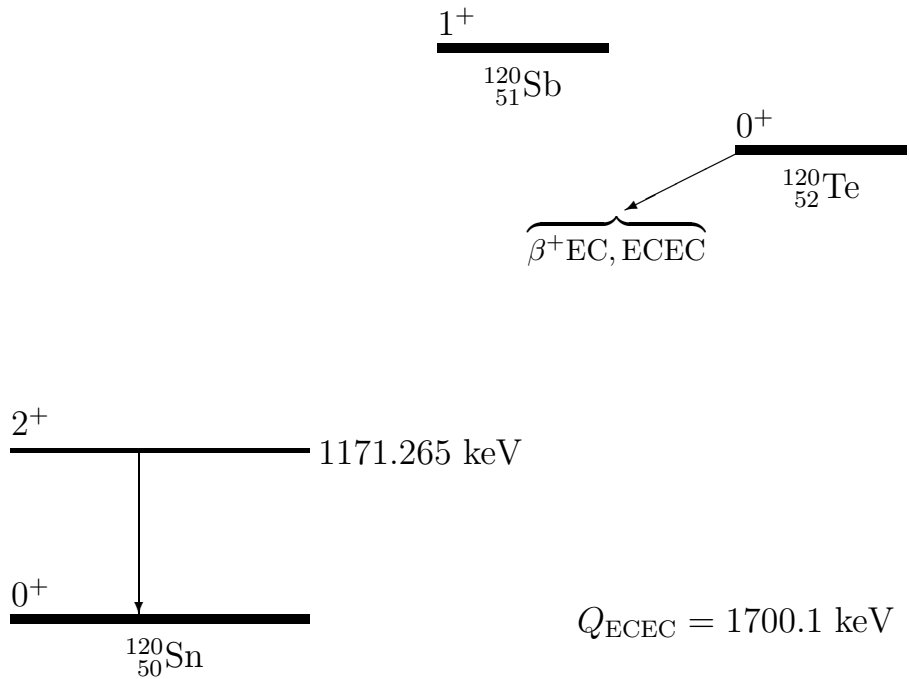}
\caption{Decay scheme of $^{120}$Te.}  
\label{fig_1}
\end{center}
\end{figure*}

\section{Experimental Details}

The experiment was performed in the Modane Underground Laboratory at a 
depth of 4800 m w.e.. The natural tellurium oxide powder sample was measured 
using a 400 cm$^3$ low-background HPGe detector.
     
The Ge spectrometer is composed of a coaxial, p-type crystal. For the 
cryostat, the main parts, i.e., the endcap and the crystal holding are made of a very 
pure Al-Si alloy. The cryostat had a J-type geometry to shield the crystal from 
radioactive impurities in the dewar and the preamplifier. The passive shielding consists of 4 cm of 
roman lead and 10 cm of OFHC copper inside 15 cm of ordinary lead. To remove
any effect of the 
$^{222}$Rn gas, one of the main sources of the background, a special effort is made 
to minimize the free space near the detector. In addition, the passive shielding is 
enclosed in an aluminum box flushed with high purity nitrogen.
     
The electronics consisted of spectrometric amplifier and an 8192 
channel ADC. The energy calibration was adjusted to cover the energy range from 50 keV 
to 3.5 MeV, 
and the resolution was 2.0 keV for the 1332-keV line of $^{60}$Co. The 
electronics were stable during the experiment due to the constant conditions in the 
laboratory (temperature of $\approx 23^\circ$ C, hygrometric degree of $\approx 50$\%).  
A daily check on the 
apparatus assured that the counting rate was statistically constant.
     
The sample of natural TeO$_2$ powder was placed in a delrin Marinelli box surrounding the 
HPGe detector. The mass of the powder was 1004.2 g, 0.72 g of which was $^{120}$Te.
Thus $3.4\times 10^{21}$ nuclei of $^{120}$Te were 
under investigation.
The duration of the measurement was 475.4 hours.
     
The search for different $\beta^+$EC and ECEC processes in $^{120}$Te 
looked for $\gamma$-ray lines corresponding to these 
processes. Introduced here is the notation $Q'$ which is the effective $Q$-value defined as 
$Q'=\Delta {\rm M} - \epsilon_1-\epsilon_2$ where $\Delta M$ is the difference of 
the parent and daughter atomic masses and $\epsilon_i$ is an electron binding energy in a daughter nuclide.
For $^{120}$Sn $\epsilon$ is equal to 29.2 keV for the K shell and 4.46 keV, 4.15 keV and 3.93 keV 
for the L shell (2s, 2p$_{1/2}$ and 2p$_{3/2}$ levels) . In the case  of the L shell 
the resolution of the detector prohibits separation of the lines so we center the study on the 4.15 keV 
line.

The ECEC$(0\nu)$ transition to the ground state of the daughter nuclei was considered for three 
different electron capture cases:

1) Two electrons are captured from the the L shell. In this case, $Q'$ is equal to $(1691.8 \pm 10.3)$ keV and 
the transition is accompanied by a bremsstrahlung $\gamma$-quantum with an energy $\sim$ 1691.8 keV.

2) One electron is captured from the K shell, another from the L shell. In this case, $Q'$ 
is equal to $(1666.8 \pm 10.3)$ keV and the transition is accompanied by a bremsstrahlung $\gamma$-quantum 
with an energy $\sim$ 1666.8 keV.  

3) Two electrons are captured from the K shell. In this case, $Q'$ is equal to $(1641.7 \pm 10.3$) keV and 
the transition is accompanied by $\gamma$-quantum with an energy $\sim$ 1641.7 keV. In fact this 
transition is strongly suppressed (forbidden) because of momentum conservation. So in this case the more probable outcome is 
the irradiation of $e^+e^-$ pair \cite{DOI93} which gives two annihilation $\gamma$-quanta with energy 511 keV.

The ECEC$(0\nu + 2\nu)$ transition, to the 2$^+$ excited state, is accompanied with $\gamma$-quanta with an 
energy of 1171.26 keV.
The $\beta^+$EC$(0\nu + 2\nu)$ transition to the ground state is accompanied by two annihilation 
$\gamma$-quanta with energy 511 keV.

Gamma-ray spectra of selected energy ranges are shown in Fig.~(2-4).
These spectra correspond to regions of interest for the different decay modes of 
$^{120}$Te.  The spectra are presented with the background from measurements without the TeO$_2$ powder
and corrected for the measuring time of the experiment. In these spectra all visible peaks in the experimental and 
background spectra are within one sigma error, so there are no extra events 
that are statistically significant at the $3\sigma$ level.

The Bayesian approach \cite{PDG04} was used to estimate limits on transitions of $^{120}$Te to the 
ground and excited states of $^{120}$Sn. To construct the likelihood function every bin of the 
spectrum is supposed to have a Poisson distribution with its mean $\mu_i$ and the number of events 
equal to the content of this $i$th bin. The mean $\mu_i$ can be written in general form as

\begin{equation}
\mu_i = N\sum_{m} {\varepsilon_m a_{mi}} + \sum_{k}
{P_k a_{ki}} + b_i
\end{equation}

The first term in (5) describes the contribution of the investigated process that may have a few 
$\gamma$-lines contributing appreciably to the $i$th bin; the parameter $N$ is the number of 
decays, $\varepsilon_m$ is the detection efficiency of the $m$th $\gamma$-line of this transition 
and $a_{mi}$ is the contribution of $m$th line to the $i$th bin. For low-background measurements a 
$\gamma$-line may be taken to have a gaussian shape. The second term gives contributions of 
background $\gamma$-lines; here $P_k$ is the area of the $k$th $\gamma$-line and $a_{ki}$ is its 
contribution to the $i$th bin. The third term represents the so-called ``continuous background'' 
$b_i$ which has been selected as a straight-line fit after rejecting all peaks in the region of 
interest.  We have selected this region as the peak investigated $\pm$ 30 standard deviations 
($\approx$ 20 keV). The likelihood function is the product of probabilities for selected bins.  
Normalizing this to 1 over parameter $N$, it becomes the probability density function for $N$, 
which is used to calculate limits for $N$.  To take into account errors of $\gamma$-line shape 
parameters, peak areas, and other factors, one should multiply the likelihood function by the 
error probability distributions for those values and integrate, to provide the average 
probability density function for $N$.

In the case of ECEC$(0\nu)$ transition to the ground state of $^{120}$Sn 
there is a large uncertainty in the energy of the bremsstrahlung $\gamma$-quantum because of poor accuracy 
for $\Delta M$ ($\pm 10.3$ keV). Thus the position of the peak was varied in the region of the 
uncertainty and the most 
conservative value of the limit for the half-life was selected.       

The photon detection efficiency for each investigated process has been computed with the CERN 
Monte Carlo code GEANT 3.21. 
Special calibration measurements with radioactive sources and powders containing well-known 
$^{226}$Ra activities confirmed that the accuracy of these efficiencies is about 10\%.

\begin{figure*}[h]
\begin{center}
\includegraphics[width=10cm]{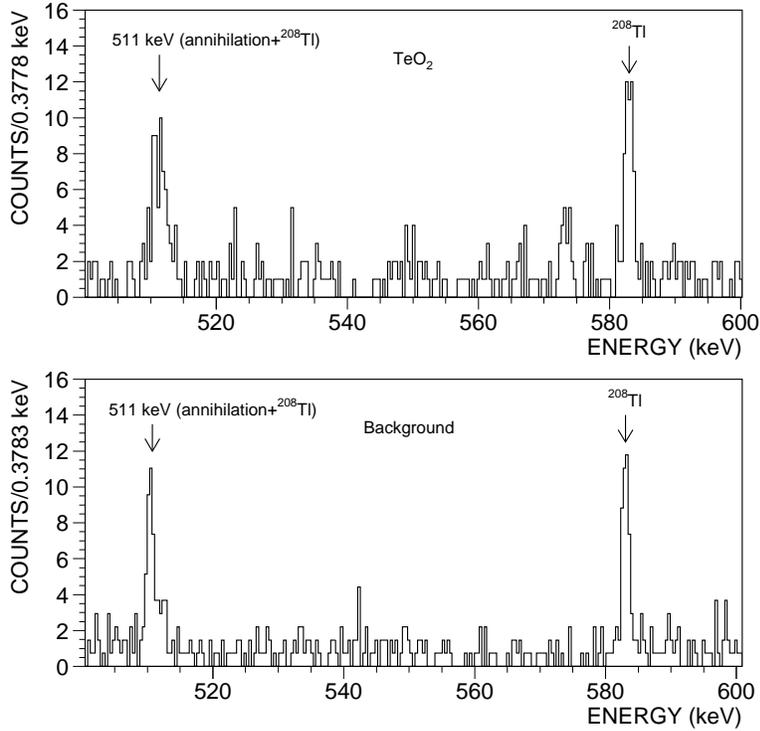}
\caption{a) Total $\gamma$-ray spectrum in the energy range of 500 to 600 keV. \\
b) Background spectrum reduced to the same time of measurement.}  
\label{fig_1}
\end{center}
\end{figure*}

\begin{figure*}[h]
\begin{center}
\includegraphics[width=10cm]{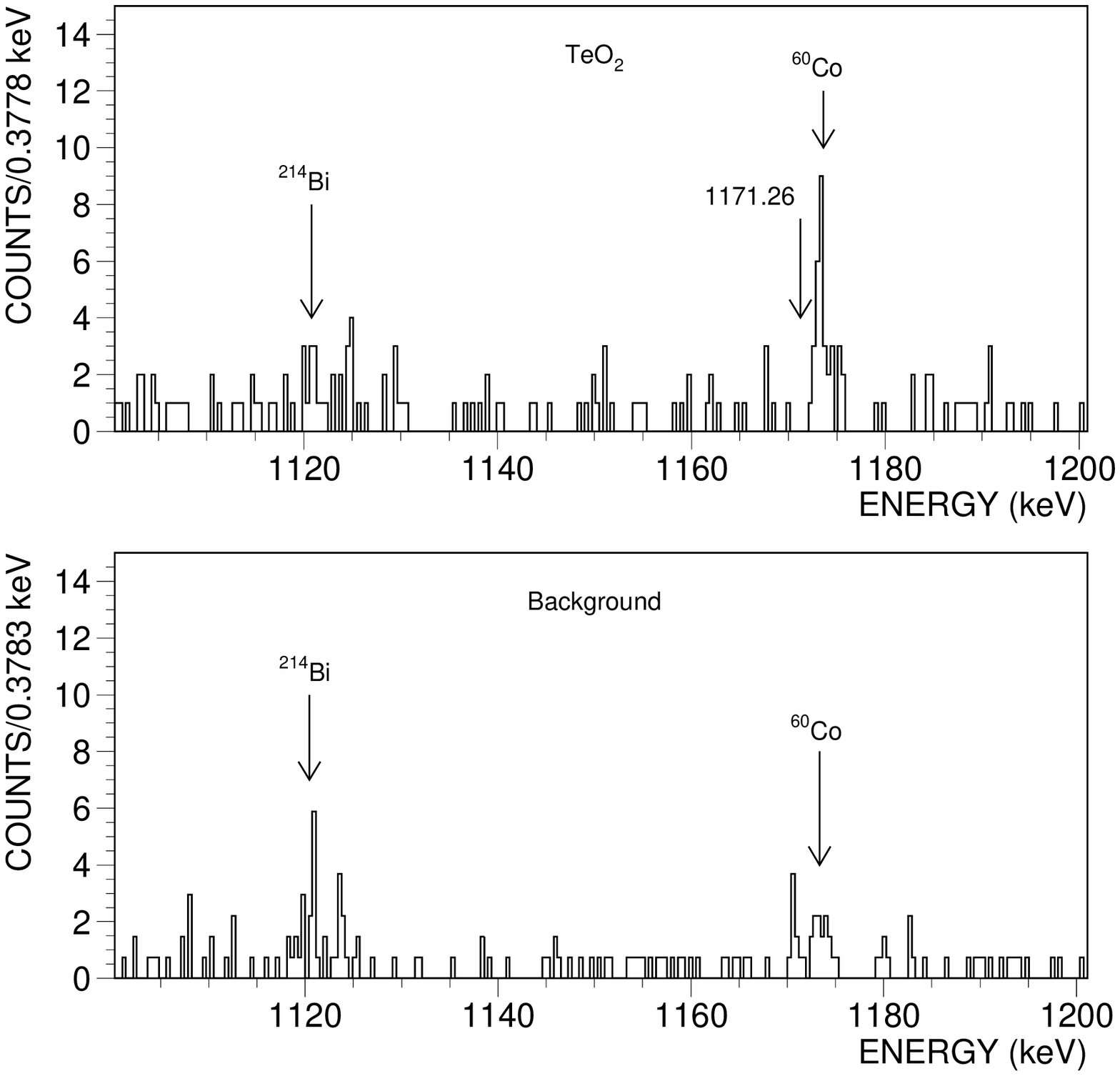}
\caption{a) Total $\gamma$-ray spectrum in the energy range of 1100 to 1200 keV. \\
b) Background spectrum reduced to the same time of measurement.}  
\label{fig_1}
\end{center}
\end{figure*}

\begin{figure*}
\begin{center}
\includegraphics[width=10cm]{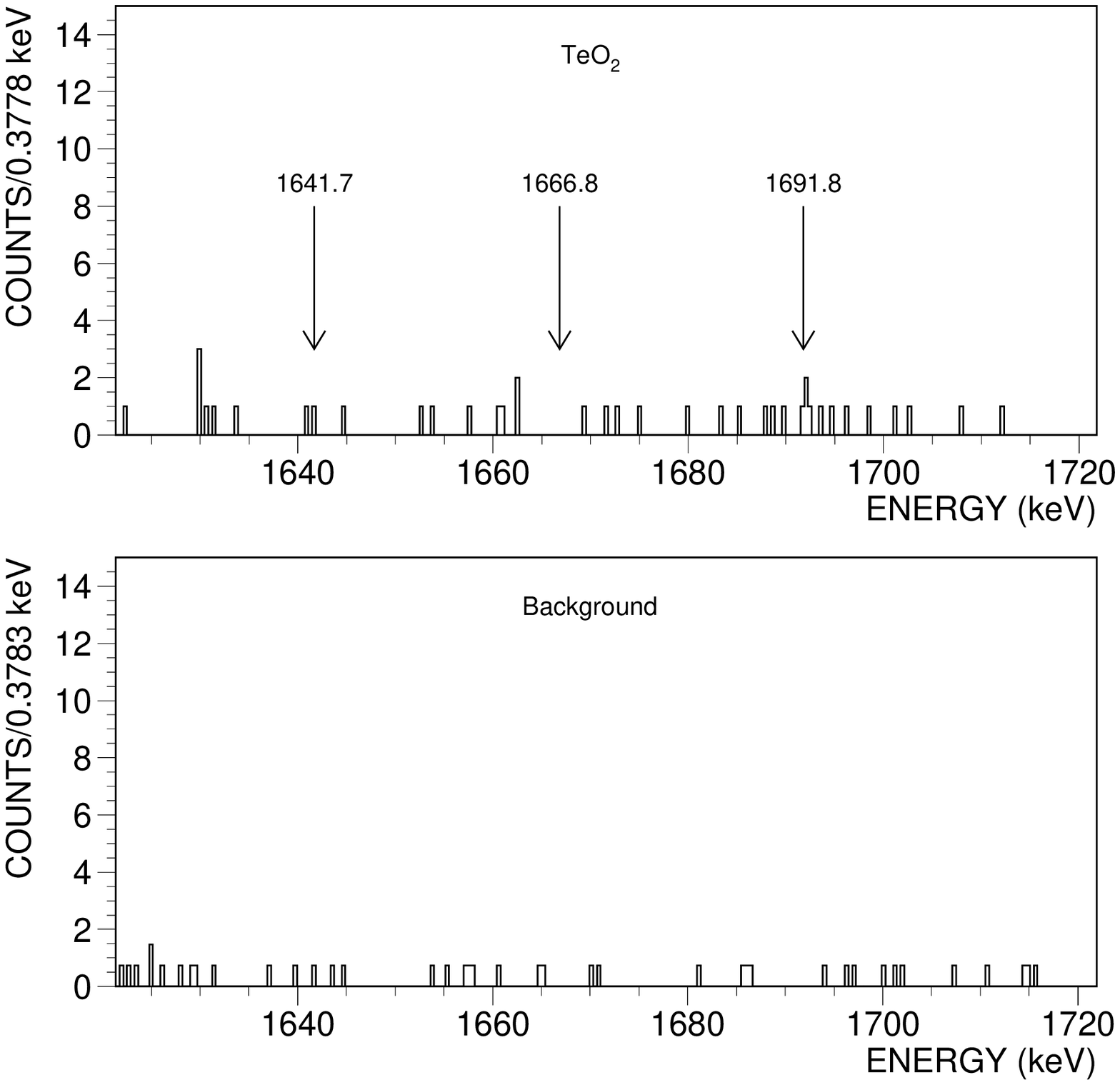}
\caption{a) Total $\gamma$-ray spectrum in the energy range of 1620 to 1720 keV. \\
b) Background spectrum reduced to the same time of measurement.}  
\label{fig_1}
\end{center}
\end{figure*}

\section{Discussion and conclusion} 

\begin{table}
\caption{\label{arttype}The limits on $\beta^+$EC and ECEC processes in $^{120}$Te. 
(See text for the details.)}
\footnotesize\rm
\begin{tabular*}{\textwidth}{llll}
\br
Transition & Energy of $\gamma$-rays (efficiency) & \multicolumn{2}{c}{$T_{1/2}$, $10^{18}$ y (C.L. 90\%)}\\
\cline{3-4}
 & & This work & other works \\
\mr
$\beta^+$EC$(0\nu + 2\nu)$; g.s. & 511.0 (7.38 \%) & 0.19 & 0.049 (0$\nu$) \cite{WIL06}\\
ECEC$(0\nu)$ ${\rm L}^1{\rm L}^2$; g.s. & 1691.2 (2.05 \%) & 0.29 & 0.0019 (0$\nu$) \cite{WIL06}\\
ECEC$(0\nu)$ ${\rm K}^1{\rm L}^2$; g.s. & 1666.4 (2.08 \%) & 0.39 & 0.0019 (0$\nu$) \cite{WIL06}\\
ECEC$(0\nu)$ ${\rm K}^1{\rm K}^2$; g.s. & 1641.7 (2.08 \%) & 0.60 & 0.0019 (0$\nu$) \cite{WIL06}\\
 & 511.0 (7.38 \%) & 0.19 & -\\
ECEC$(2\nu)$; g.s. & - & - & 0.0094 (2$\nu$) \cite{KIE03}\\
ECEC$(0\nu + 2\nu)$; 2$^+$ & 1171.26 (2.60 \%) & 0.75 & 0.0084 (2$\nu$) \cite{KIE03}\\

\br
\end{tabular*}
\end{table}

The final results are presented in Table 1. In the last column the best previous results 
are presented for comparison. One can see that the present work is 4 to 300 times better 
than previous measurements \cite{KIE03,WIL06}.
The experimental 
limits are quite interesting as they exclude some unexpected processes.
Our approach is not sensitive to the ECEC$(2\nu)$ capture to the ground state because X-rays are 
absorbed in the sample and can not reach the sensitive volume of the HPGe detector.

The isotope $^{120}$Te has only minimally been investigated theoretically. The only paper \cite{ABA84} 
gives half-life values for the $\beta^+EC(2\nu$) and ECEC(2$\nu$) processes which 
were estimated by assuming the dominance of the lowest 1$^+$ state of the intermediate isobar. 
The theoretical values from this reference are 
$4.4\times 10^{26}$ y and $1.8\times 10^{24}$ y respectively. Unfortunately there is no experimental 
information about $\beta^-$decay of $^{120}$Sb and this fact degrades the quality of the estimations. 
Thus, there are no really good predictions with which to 
compare our results. Nevertheless, we will try to estimate the significance of our results and 
the possibility to increasing the sensitivity of this type of experiment in the future.

The experimental possibilities are the following:\\
1) With 1 kg of enriched $^{120}$Te in the setup described earlier, 
the sensitivity after one year of measurement will be $\sim 3\cdot 10^{21}$ y. \\
2) Different $\beta^+EC$ and ECEC processes in $^{120}$Te could be investigated in the 
current CUORICINO \cite{ARN05} experiment
where $\sim$ 40 kg of natural Te are under investigation. The ECEC$(2\nu)$ transition to the ground state 
of $^{120}$Sn could be investigated as well. The sensitivity for the above processes  
is estimated as $\sim$ $10^{21}$ y.\\ 
3) The future CUORE \cite{ARN04} experiment with 1000 t of TeO$_2$ will have a sensitivity of 
up to $\sim$ $10^{24}$ y. Approximately the same level of sensitivity can be achieved in 
future COBRA  experiment \cite{WIL06} given that a very low background level is reached.
One can estimate the half-life for $^{120}$Te from the results of \cite{ABA84} or the half-life 
estimations for these processes in 
different nuclei with close transition energies \cite{DOM05,RAI06}. In summary the estimates suggest that 
the sensitivity to detect the ECEC($2\nu$) process in $^{120}$Te could be reached in these future experiments.

To conclude, in this paper we have presented new, more stringent, experimental limits for 
the different $\beta^+$EC and ECEC processes in $^{120}$Te. The experimental 
limits are quite interesting as they prohibit some unexpected processes. The potential for further investigation 
of these processes in future experiments have been demonstrated.

\section*{Acknowledgments}

The authors would like to thank the Modane Underground Laboratory staff for their technical 
assistance in running experiment. The authors are very thankful to Prof. S. Sutton (MHC, USA) 
for his useful corrections and remarks. This work was supported by Russian Federal Agency for Atomic Energy.

\end{document}